\documentclass[apj]{emulateapj}
\usepackage{apjfonts}
\usepackage{amsbsy}

\usepackage{color}
\usepackage{rotating}

\def\wp{$w_p(r_p)$}
\def\hmpc{$h^{-1}$Mpc}

\def\msol{M$_\odot$}
\def\hmsol{$h^{-1}$M$_\odot$}

\def\om{\Omega_m}

\def\s8{\sigma_8}

\def\lcdm{$\Lambda$CDM}

\def\x2{$\chi^2$}
\def\hmsol{$h^{-1}\,$M$_\odot$}

\def\NNm1{\langle N(N-1) \rangle}

\def\fsat{f_{\rm sat}}
\def\slogm{\sigma_{{\rm log}M}}
\def\m_star{M_\ast}

\def\lcdm{$\Lambda$CDM}
\def\slogm{\sigma_{{\rm log}M}}
\def\om{\Omega_m}

\def\s8{\sigma_8}

\def\hmpc{$h^{-1}\,$Mpc}

\def\x2{$\chi^2$}
\def\hmsol{$h^{-1}\,$M$_\odot$}

\def\wp{$w_p(r_p)$}

\def\NNm1{\langle N(N-1) \rangle}

\def\fsat{f_{\rm sat}}

\def\fsat{f_{\rm sat}}

\def\p0{P_0(r)}

\def\msol{$M$_\odot}
\def\mgal{M_{\ast}}
\def\msol{M_\odot}
\def\vmax{V_{\rm max}}
\def\wise{WISE}
\def\deg2{deg$^2$}
\def\wp{w_p(r_p)}
\def\kpch{kpc/$h$}
\def\slogm{\sigma_{\log M\ast}}
\def\mhalo{M_{\rm halo}}

\def\smf{\Phi(\log M_{\ast})}

\begin{document}

\title{The Correlation Between Halo Mass and Stellar Mass\\ for the Most
Massive Galaxies in the Universe}

\author{ Jeremy L. Tinker$^1$, Joel R. Brownstein$^2$, Hong Guo$^3$, Alexie
  Leauthaud$^4$, Claudia Maraston$^5$, Karen Masters$^5$, Antonio
  D. Montero-Dorta$^2$, Daniel Thomas$^5$, Rita Tojeiro$^6$,
  Benjamin Weiner$^7$, Idit Zehavi$^8$, Matthew D. Olmstead$^9$}
\affil{$^1$Center for Cosmology and Particle Physics, Department of
  Physics, New York University, New York, NY 10013, USA}
\affil{$^2$Department of Physics and Astronomy, University of Utah, Salt
  Lake City, UT 84112, USA}
\affil{$^3$Key Laboratory for Research in Galaxies and Cosmology, Shanghai Astronomical Observatory, Shanghai 200030, China}
\affil{$^4$Kavli IPMU (WPI), UTIAS, The University of Tokyo, Kashiwa, Chiba 277-8583, Japan}
\affil{$^5$ICG-University of Portsmouth, PO13FX Portsmouth, UK}
\affil{$^6$School of Physics and Astronomy, University of St Andrews, St Andrews, KY16 9SS, UK}
\affil{$^7$Steward Observatory, 933 N. Cherry St., University of Arizona,
  Tucson, AZ 85721, USA}
\affil{$^8$ Department of Astronomy \& CERCA, Case Western Reserve University, 10900 Euclid Avenue, Cleveland, OH 44106, USA}
\affil{$^9$Department of Chemistry and Physics, King's College, 133 North River St, Wilkes Barre, PA 18711, USA}

\begin{abstract}

  We present measurements of the clustering of galaxies as a function
  of their stellar mass in the Baryon Oscillation Spectroscopic
  Survey. We compare the clustering of samples using 12 different
  methods for estimating stellar mass, isolating the method that has
  the smallest scatter at fixed halo mass. In this test, the stellar
  mass estimate with the smallest errors yields the highest amplitude
  of clustering at fixed number density. We find that the PCA stellar
  masses of \cite{chen_etal:12} clearly have the tightest correlation
  with halo mass. The PCA masses use the full galaxy spectrum,
  differentiating them from other estimates that only use optical
  photometric information.  Using the PCA masses, we measure the
  large-scale bias as a function of $\mgal$ for galaxies with
  $\log\mgal\ge 11.4$, correcting for incompleteness at the low-mass
  end of our measurements. Using the abundance-matching ansatz to
  connect dark matter halo mass to stellar mass, we construct
  theoretical models of $b(\mgal)$ that match the same stellar mass
  function but have different amounts of scatter in stellar mass at
  fixed halo mass, $\slogm$. Using this approach, we find
  $\slogm=0.18^{+0.01}_{-0.02}$. This value includes both intrinsic
  scatter as well as random errors in the stellar masses. To partially
  remove the latter, we use repeated spectra to estimate statistical
  errors on the stellar masses, yielding an upper limit to the
  intrinsic scatter of 0.16 dex.

\end{abstract}


\section{Introduction}

Galaxies are born, live, and die, within dark matter halos. We have
convincing evidence that the evolutionary history of galaxies and
halos is correlated to a strong degree: brighter, bigger galaxies have
higher clustering, indicative of being in more massive dark matter
halos (see, e.g., \citealt{norberg_etal:02, zehavi_etal:05,
  zehavi_etal:11} for analyses at $z\sim 0$ and \citealt{coil_etal:06,
  zheng_etal:07, wake_etal:11, leauthaud_etal:12_shmr} at higher
redshifts). Thus the growth of galaxies is related to the growth of
dark matter halos. But how correlated are these two quantities? The
purpose of this paper is to quantify this correlation by constraining
the scatter in stellar mass at fixed halo mass for galaxies in the
Baryon Oscillation Spectroscopic Survey (BOSS;
\citealt{dawson_etal:13}). We will use two-point clustering as our
probe of this scatter, $\slogm$. For massive galaxies, clustering is
an especially sensitive diagnostic of the scatter because it directly
impacts their large-scale bias (e.g., \citealt{reddick_etal:13}); more
scatter means a sample of galaxies will contain a more significant
sample of low-mass halos that will lower the overall clustering
amplitude. BOSS galaxies specifically represent an excellent sample of
highly biased objecs, with previous small-scale measurements yielding
clustering amplitudes roughly four times higher than that of dark
matter (\citealt{white_etal:11, guo_etal:13, saito_etal:16}).

However, unlike magnitude and color, galaxy stellar mass is not an
observable. Different methods for deriving $\mgal$ produce different
results. The scatter we constrain through clustering is the quadrature
sum of the intrinsic scatter of stellar mass at fixed halos,
$\sigma_{\rm int}$ and and measurement error $\sigma_{\rm err}$.
Different methods of deriving stellar mass will have different
$\sigma_{\rm err}$ but have the same intrinsic scatter since they are
all estimates of the same physical quantity. Thus, we can also use
clustering to determine which method of obtaining stellar mass has the
least variance. In this context, systematic offsets between codes or
choices within codes, such as switching stellar initial mass
functions, are immaterial. What we care about here is the
rank-ordering of galaxies from most massive to least massive. Using
clustering, we cannot probe systematic offsets between stellar mass
estimates.

In addition to comparing different methods for determining stellar
mass, we compare stellar mass to other physical properties of the
galaxy. Specifically, \cite{wake_etal:12} used clustering to claim
that stellar velocity dispersion, $\sigma_{\rm vel}$, correlates
better with halo mass than $\mgal$. We will compare the clustering of
galaxies ranked by 12 different variations of thee different stellar
mass codes, all of which are available in the public SDSS data
releases, as well as two estimates of $\sigma_{\rm vel}$ and multiple
galaxy luminosities.

For converting redshift to distance, as well as calculating $\mgal$,
we assume a flat \lcdm\ cosmology with $\om=0.3$ and $h=0.7$. 

\section{Data}

We use results from Data Release Ten of the Sloan Digital Sky Survey
(\citealt{dr10}). The spectroscopic footprint covers 6895 deg$^2$
combined in the North Galactic Cap and South Galactic Cap regions,
roughly 70\% of the full BOSS footprint. 

\subsection{The CMASS Sample}

The CMASS target sample is the workhorse of the BOSS large scale
structure analysis. Further details can be found in
\cite{dawson_etal:13}. These color cuts are meant to isolate massive
galaxies at $z\gtrsim 0.4$, but are somewhat more inclusive of blue
galaxies than the traditional luminous red galaxy (LRG) sample from
SDSS-II (\citealt{eisenstein_etal:01}). To limit the effects of the
color cuts and flux limit of the survey, our fiducial results are
restricted to the range $z=[0.45, 0.60]$, which surrounds the peak in
the redshift distribution around $4\times 10^{-4}$ (\hmpc)$^3$.

Figure \ref{color_mass} shows the mass-vs-$g-i$ color distribution for
CMASS galaxies. We will describe the mass estimate used in this figure
(PCA) in \S \ref{s.mass_estimates}. We use $g-i$ because it more
naturally separates blue and red galaxies at these masses and
redshifts than other colors. \cite{masters_etal:11} demonstrates that
a simple $g-i>2.35$ color cut best separates early type and late type
CMASS galaxies. At these redshifts, $g-i=2.35$ is similar to
$u-r=2.22$ at $z=0$, used by \cite{strateva_etal:01} to describe
bimodality in the local universe. Unlike samples that probe galaxies
near the knee in the stellar mass function (e.g.,
\citealt{blanton_etal:03cmd}), the CMASS sample is not bimodal in its
colors. Although the color cuts are specifically designed to be more
efficient at selecting passive objects than actively star-forming
objects, the dominant factor in the lack of bimodality is simply that
the CMASS selection is targeting the very massive end of the galaxy
distribution, with a median stellar mass of $\mgal\sim
10^{11.5}\,\msol$. At these masses, even the SDSS Main sample
exhibits no bimodality, but rather has a tail to bluer colors
comprised of the few star-forming objects at these mass scales. The
CMASS selection is not devoid of any star-forming objects;
\cite{chen_etal:12} find roughly $3\%$ of CMASS objects at the median
stellar mass have formed 10\% of their stellar mass in the last Gyr.

\subsection{The SPARSE Sample}

The CMASS\_SPARSE sample (hereafter SPARSE for brevity) targets the
same luminosity range but with a slightly expanded color range. The
density of target galaxies is sensitive to the exact value of the
color-magnitude intercept, thus the SPARSE sample is meant to expand
the color range of the CMASS in order to test any biases in the
selection and to probe to lower stellar masses. The median redshift of
the SPARSE sample is notably lower than CMASS due to the wider color
cuts. Although the color shift is only 0.28 magnitudes, there are
nearly as many objects that pass the SPARSE selection (and are not
included in the CMASS sample) as in the CMASS sample itself. The name
SPARSE is derived from the fact that these galaxies are randomly
subsampled by a factor of 5 (to 5 targets/deg$^2$) in order to
minimize the number of fibers allocated to this sample (see
\citealt{dawson_etal:13} for more details on the target class as
well).

\begin{figure}
\epsscale{1.2} 
\plotone{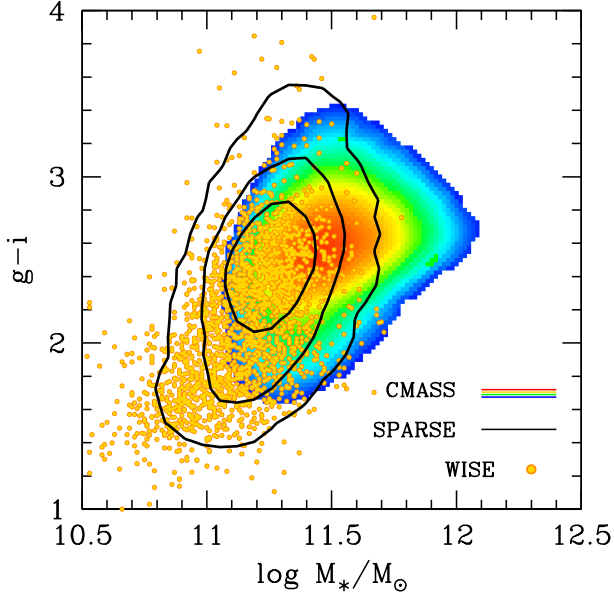}
\vspace{-0.5cm}
\caption{ \label{color_mass} The $g-i$ color-mass distributions of the
  three different target classes utilized in this paper. The color
  contours represent the CMASS galaxy sample, with color representing
  density of points in each cell. The black curves represent the
  density contours of SPARSE targets. The WISE targets are sparse
  enough that they can be shown individually, represented by the filled
  circles. All results are shown in the redshift range of $z=[0.45,
  0.60]$.}
\end{figure}

Figure \ref{color_mass} shows the color-mass distribution of these
targets in comparison to the CMASS to sample. After trimming the
SPARSE catalog to objects within $z=[0.45,0.60]$, the stellar
masses probed by the SPARSE selection shifts the median stellar mass
lower by $\sim 0.2$ dex, and the mode of the $g-i$ color distribution
also shifts blueward by $\sim 0.2$ magnitudes. But, as with CMASS, the
SPARSE sample is dominated by passive red galaxies and does not
exhibit any bimodality, although the blue tail extends lower in $g-i$
than the CMASS sample.

The definition of the SPARSE sample has fluctuated somewhat over the
course of the survey. In the first few months of observations, the
color range was somewhat broader. We exclude these areas from
consideration in the statistics, removing $\sim 140$ deg$^2$ of the
overall footprint (delineated ``chunk2'' in the BOSS order of
observations; see \citealt{dawson_etal:13}).

\subsection{The WISE\_COMPLETE Sample}

Relaxing the color cuts even further than the SPARSE sample would
dramatically reduce the efficiency of finding $z\gtrsim 0.4$
galaxies. To efficiently locate galaxies in our desired redshift
range, but outside the BOSS color cuts, ancillary data must be brought
in. To this end, a series of 26 plates covering 59.8 deg$^2$ were
dedicated to observing three sets of ancillary targets, one of which
was the WISE\_COMPLETE target set. These targets incorporate data from
the Wide-field Infrared Survey Explorer (WISE; \citealt{wise}). These
targets have the same magnitude range as CMASS and SPARSE, but employ
a single color cut,

\begin{equation}
\label{e.wise}
r-W1>4.165,
\end{equation}

\noindent to select galaxies in our desired stellar mass and redshift
range that are outside the standard BOSS color cuts\footnote{In the
  input catalog, we removed any galaxy that also passed the CMASS
  cuts. In post-processing, we removed any target that also passed the
  SPARSE selection. This latter cut encompassed 11\% of the input
  \wise\ catalog.}. Here, $W1$ is the WISE 3.4 micron band. Using WISE
near-IR photometry is promising because the IR bands are on the far
side of the peak of stellar emission, so an optical-IR color is
sensitive to redshift. We will refer to the WISE\_COMPLETE sample as \wise\
for brevity.

There are 7368 objects within the \wise\ target catalog that received
fibers and recovered accurate redshifts, 2200 of which are within the
redshift range of interest. The $g-i$ color distribution is broad and
flat, with a median stellar mass of $\mgal=10^{11.06} \msol$.

\begin{figure}
\epsscale{1.2} 
\plotone{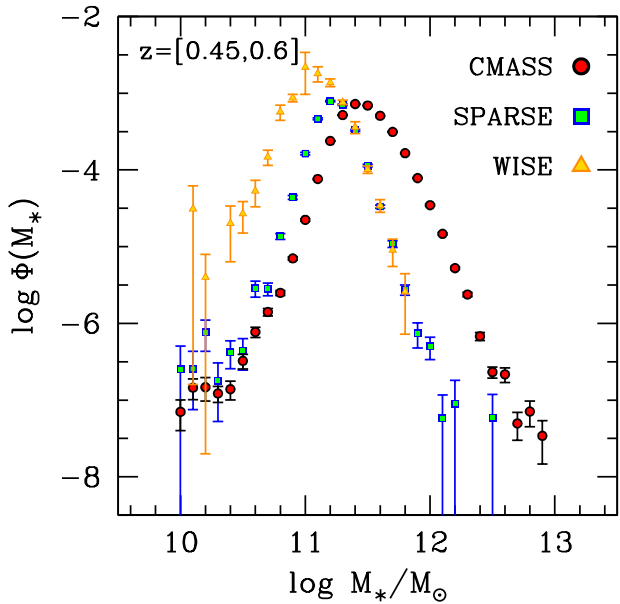}
\vspace{-0.5cm}
\caption{ \label{smf_compare} Comparison of the space densities of
  each individual BOSS targets class. All measurements implement at
  $\vmax$ correction, which has minimal effect at $\log\mgal\ge 11.4$
  but can significantly enhance the measured abundance at low
  masses. Red circles represent the CMASS targets. Green squares
  represent the SPARSE sample. Yellow triangles represent the \wise\
  data. Each of these measurements peaks at a lower stellar mass,
  demonstrating the effect of reaching further outside the typical LRG
  color selection.}
\end{figure}

% \subsection{BOSS data} --> this was removed

\subsection{Physical properties of the BOSS galaxies}
\label{s.mass_estimates}

As discussed in the introduction, we are interested in finding the
physical property that correlates the closest with halo mass via the
clustering of the sample. We use stellar mass, galaxy luminosity, and
galaxy velocity dispersion. Our main focus is stellar mass, but the
question of whether stellar mass correlates any better with halo mass
than does luminosity is still open. Table \ref{t.props} shows all 16
variations of galaxy properties we use to rank-order the CMASS galaxy
sample. The first 12 are stellar masses that are included in the
standard BOSS pipeline. There are three different codes for deriving
stellar masses from BOSS data, with each code producing variants based
upon parameter choices (i.e., stellar initial mass function, stellar
population synthesis code, etc.).

The Portsmouth masses (\citealt{maraston_etal:13}) use two template
star formation histories, one to describe a passively evolving, old
stellar population meant to mimic an LRG-type galaxy, and the other is
an actively star-forming template. These two templates are combined to
make one catalog: each galaxy is assigned a template based on its
color, with a redshift-dependent break point of $g-i=3.25
+1.67(z-0.53)$. Galaxies redder than this color are assigned the
passive template, which bluer galaxies are assigned the star-forming
template. Thus, although we list 14 different stellar mass definitions
in Table \ref{t.props}, there are only 12 different stellar mass
catalogs, two of which are comprised of the Portsmouth masses.  The
Granada masses are an implementation of the Flexible Stellar
Population Synthesis (FSPS) mass modeling described in
\cite{conroy_etal:09} . The PCA masses (`principal component
analysis') are described in \cite{chen_etal:12}. In contrast to the
first two methods, the PCA approach uses the entire galaxy spectrum,
decomposing it into principal components which are then fit to a
library of varying galaxy templates to obtain the physical properties
of the galaxies, including $M/L$ ratio, velocity dispersion, and star
formation rate.

We use two different estimates of the stellar velocity dispersion, one
that is produced by the PCA analysis, the other produced by the
Portsmouth analysis code (see details for the latter in
\citealt{thomas_etal:13}).. Lastly, we use absolute $i$-band magnitude
as our baselines rank-order by which to compare all the methods. We
also include a sample that has been $k$-corrected to $z=0.52$ using
the {\tt kcorrect} code of \cite{blanton_roweis:07}.

\section{Measurements}

\subsection{The Completeness and Abundance of the BOSS Galaxy Samples}

Figure \ref{smf_compare} shows the stellar mass function, $\smf$ for
each of the three samples discussed above. All results use the PCA
stellar masses. We estimate $\smf$ using the $1/V_{\rm max}$ method,
while noting that the $V_{\rm max}$ weighting really only effects the
results at stellar masses below the peak of each distribution. The
main source of incompleteness in these measurements is the color
selection imposed on each sample. The CMASS sample dominates the
statistics at high masses, but the combination of the SPARSE and
\wise\ samples becomes equivalent to the CMASS abundance at
$\log\mgal\sim 11.4$.

\begin{figure}
\epsscale{1.2} 
\plotone{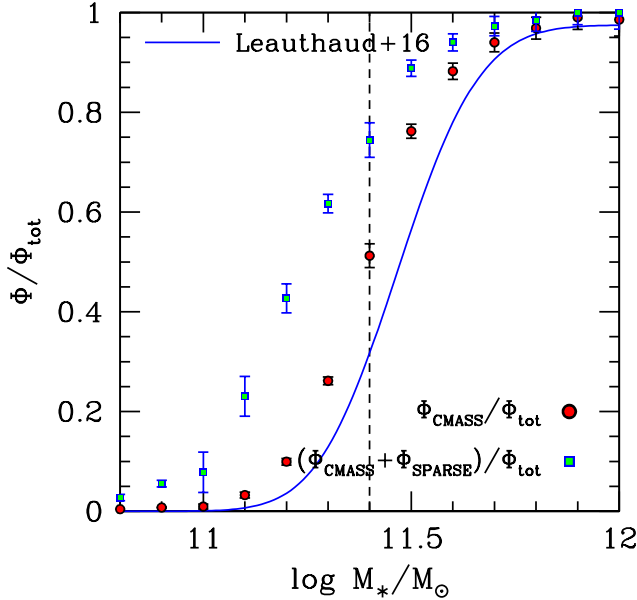}
\vspace{-0.5cm}
\caption{ \label{completeness} Completeness of the BOSS target
  samples, relative to the total abundance of all three target
  classes. At $\log\mgal=11.4$, indicated by the vertical dash line,
  the CMASS sample is only 50\% complete. Adding the SPARSE sample
  brings the completeness at this mass scale up to 75\%. The CMASS
  sample by itself is >95\% complete at $\log\mgal>11.7$. At lower
  masses, the total abundance is itself incomplete. We thus make a
  conservative completeness limit of $\log\mgal=11.4$ for the full
  BOSS sample, at which scale both SPARSE and \wise\ abundances have
  steep mass functions. }
\end{figure}

We use the abundance of SPARSE and \wise\ galaxies to quantify the
completeness of the BOSS target selection. Figure \ref{completeness}
shows the ratio of the CMASS SMF to that of the total SMF (i.e.,
CMASS+SPARSE+\wise) as a function of $\log\mgal$. The CMASS sample is
roughly 50\% complete in stellar mass at $\log\mgal=11.4$. These
results are in reasonable agreement with those of
\cite{leauthaud_etal:16}, which uses the extra near-infrared imaging
data available in Stripe 82 to estimate completeness. Although the
combination of WISE and SPARSE data get us most of the way to a fully
complete sample, the total sample used in this paper is still missing
$\sim 10\%$ of galaxies at $\log M=11.4$. The stellar masses of
\cite{leauthaud_etal:16} are different than the PCA masses used here
due to the inclusion of infrared imaging data. From the comparison
between the near-IR masses and the PCA \cite{bundy_etal:15}, there is
a shift of 0.15 dex between the two mass definitions, which we use to
shift the completeness \cite{leauthaud_etal:16} completeness curves
onto the PCA mass scale.

At $\log\mgal = 11.4$, the combination of CMASS and SPARSE is roughly
75\% complete. Henceforth, we will limit all analyses to be above this
mass scale. The limited size of the \wise\ sample prevents us from
making robust clustering measurements for \wise\ galaxies, thus we
make the assumption that the bias of \wise\ galaxies and SPARSE
galaxies is the same when combining the clustering results of the
samples. This does not significantly affect our results; the
constraining power on $\sigma_{\log\mgal}$ derive from the region of
the SMF where CMASS dominates the statistics.

\begin{figure}
\epsscale{1.2} 
\plotone{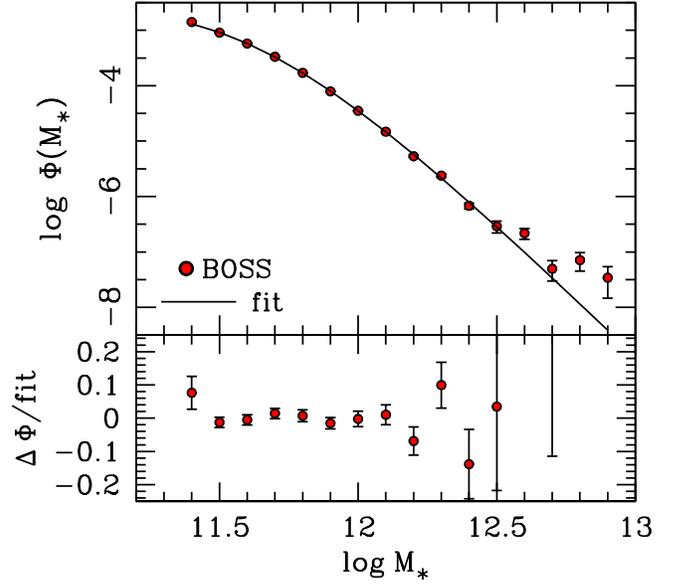}
\vspace{-0.5cm}
\caption{ \label{smf_dr10_fit} {\it Upper Panel:} The total BOSS
  stellar mass function at $z=[0.45,0.60]$, down to our completeness
  limit of $\log\mgal=11.4$. The best-fit parameters of equation (x)
  are shown with the solid curve. {\it Lower panel:} Residuals of the
  fit. The $y$-axis is $(\Phi_{BOSS}-\Phi_{fit})/\Phi_{fit}$. The
  $\chi^2$ for this fit is 7.0 for $(10-5)$ degrees of freedom.}
\end{figure}

Figure \ref{smf_dr10_fit} shows the completeness-corrected SMF of
massive galaxies down to the limit of $\log\mgal=11.4$. We fit these
data using a fitting function of the form:

%set xx = 10**(m-a2)
%set x2 = 10**(m-a7)
%set f1 = a1*xx**(1+a3)*(1+xx**a4)**((a5-a3)/a4)*ln(10)

\begin{equation}
\Phi(\mgal) = \Phi^\ast\left(\frac{\mgal}{M_1}\right)^{(1+\alpha)}
\left[1+\left(\frac{\mgal}{M_1}\right)^\beta\right]^{(\gamma-\alpha)/\beta}\ln(10)
\end{equation}

\noindent with best-fit parameter values of
$\Phi^\ast=2.43\times10^{-2}$, $\log M_1=11.47$, $\alpha=0.728$,
$\beta=1.173$, and $\gamma=-6.01$. This fitting function will be used
to map galaxy mass onto halo mass using the abundance matching method
in \S \ref{s.slogm}.

\subsection{Measuring Clustering}

Our statistical measure of the clustering of galaxies within BOSS is
the projected two-point correlation function $\wp$. This statistic
integrates the two-dimensional redshift-space correlation function,
$\xi(r_p,r_\pi)$, along the line-of-sight direction $r_\pi$. $\wp$ is
defined as

\begin{equation}
\wp = 2\times \int_0^{\pi_{\rm max}} \xi(r_p,r_\pi) dr_\pi,
\end{equation}

\noindent where $r_p$ is the projected separation between galaxy pairs
and $\pi_{\rm max}$ is the maximal line-of-sight separation, which
here we make 80 \hmpc. To estimate $\xi(r_p,r_\pi)$ we use the
Lansy-Szalay estimator (\citealt{landy_szalay:93}) with $10^7$
randoms. When $\pi_{\rm max}=\infty$, the projected correlation
function is identical whether the argument in the integrand is the
real- or redshift-space correlation function. Having a finite
$\pi_{\rm max}$ introduces some deviation from a real-space-only
calculation (see, e.g.,  \citealt{vdb_etal:13}).

The projected clustering of galaxies is our preferred statistic for
two reasons. First, peculiar motions are (mostly) removed in the
line-of-sight integration. At large scales there is still a residual
effect, but we will model this out when fitting for bias. Second, in
the third paper in this series, we will present halo occupation fits
of the clustering to determine the stellar-to-halo-mass ratio of BOSS
galaxies. Removing the peculiar motions makes such modeling less
cosmologically dependent and easier to implement analytically.

To estimate the errors on the correlation function, we jackknife the
DR10 footprint into 100 roughly equal-size angular regions, removing
one subsample at a time and calculating the covariance matrix as

\begin{equation}
C_{ij} = \frac{N-1}{N}\sum_{n=1}^N \left(w_i - \bar{w}_i\right)
\left(w_j - \bar{w}_j\right),
\end{equation}

\noindent where $N$ is the total number of jackknife subsamples (100),
and $i$ and $j$ represent $r_p$ bins, and $\bar{w}$ represents the
mean correlation function in each bin. 

To correct for fiber collisions, we use the angular-upweighting method
as described in \cite{white_etal:11}. On a given plate, fibers can
only be positioned within 62 arcsec of one another. At $z=0.5$, this
translates to a projected separation of about 400 \kpch. Roughly
40\% of the area within the survey is covered by multiple
plates. These areas have nearly unit completeness, at least in terms
of fiber assignment. Galaxy pairs within the collision radius are
upweighted by the ratio of the angular correlation function of all
CMASS targets to those for which fibers were assigned at the angle of
the pair in question. Inside 62'', this ratio\footnote{Which, incidentally, is the ratio of total area
  of the survey to that covered by more than one plate.} is nearly a constant
value of 2.57. In practice,
we only use this method to correct for one $r_p$ bin inside the
collision radius, and for no bins that effect our bias calculation.

We measure the autocorrelation of the CMASS sample, but the SPARSE
sample, by definition, has too low a density to make a robust
measurement of the clustering. We thus cross-correlate the SPARSE and
CMASS samples to get the relative bias between these two samples and
then divide by the bias of the overall CMASS sample, which is $1.99\pm
0.04$ for the cosmology chosen.

\subsection{Measuring Bias}
\label{s.measuring_bias}

We use the projected separation range $5< r_p <35$ \hmpc\ to determine
the bias relative to the clustering of matter. Although clustering is
more linear at larger scales, errors on clustering rise monotonically
with scale. Additionally, as $r_p$ increases, the contribution of
peculiar velocities to $\wp$ also monotonically increases. Although we
model out the peculiar motions, the optimal approach is to extend the
$r_p$ range out as far into the linear regime as is possible, while
keeping the effect of a finite $\pi_{\rm max}$ to under 25\%. 

To recover linear bias from our clustering measurements, we need to
model both peculiar velocities and non-linear clustering. Rather than
try to subtract these effects out of our measurements, we add them
into the model for the matter clustering. First, we model real-space
galaxy clustering as

\begin{equation}
\xi_{\rm gal} (r) = b_{\rm gal}^2\xi_m(r)\zeta(r),
\end{equation}

\noindent where $\xi_m(r)$ is the non-linear matter correlation
function from \cite{smith_etal:03}, $b_{\rm gal}$ is the bias
parameter for the galaxy sample, and $\zeta(r)$ is the scale-dependent
halo bias function from equation (B7) in \cite{tinker_etal:05}. This
scale-dependence of halo bias, relative to the large-scale value,
asymptotes to unity at $r\gtrsim 20$ \hmpc, with a maximum deviation
of $\sim 10\%$ at $r\sim 2$ \hmpc\ (see Figure 12 in
\citealt{white_etal:11} for a halo occupation fit of the
scale-dependent bias of BOSS CMASS galaxies). At smaller scales, the
scale-dependence of galaxy bias becomes dependent on the details of
halo occupation and it is only partially correlated with bias at large
scales.

Second, we incorporate the effects of peculiar velocities using the
linear approximation of \cite{kaiser:87}. Although peculiar velocities
are poorly described by linear theory at $r\lesssim 20$ \hmpc, at
$\pi_{\rm max}$ of 80 \hmpc\ eliminates the contribution of redshift
space distortions at scales where linear theory breaks down. The
configuration-space implementation of the Kaiser effect is described
in detail in Appendix A of \cite{hawkins_etal:03}.

In practice we fit for $b_{\rm gal}$ by $\chi^2$ minimization using
the full covariance matrix.

\subsection{Numerical Simulations}
\label{s.nbody}

Our theoretical models are created using the publicly available halo
catalog from the MultiDark simulation (\citealt{nuza_etal:13}). The
cosmology for this simulation is $\om=0.27$ and $\s8=0.82$. The halo
catalog incorporates halos and subhalos using the {\small ROCKSTAR}
algorithm of \cite{rockstar}. We populate these halos with galaxies
using the sub-halo abundance-matching model (a.k.a. SHAM; see
\citealt{reddick_etal:13} and references therein). In its simplest
form, abundance matching provides a unique mapping of halo mass onto
galaxy mass (or galaxy luminosity) by assuming a one-to-one relation
between the two without any scatter. There are multiple ways to
incorporate scatter (\citealt{behroozi_etal:10,
  trujillo_gomez_etal:11}). We use the method in
\cite{wetzel_white:10}, in which the stellar mass function is first
deconvolve of lognormal scatter of width $\slogm$. This allows us to
employ the simple abundance matching method using the monotonic
relation between $\mgal$ and $M_h$. Then the $\log\mgal$ of each
galaxy is shifted by a random Gaussian deviate of the same width.

Once halos are populated with galaxies, the large-scale bias of the
model is calculated by binning the mock galaxies by $\log\mgal$ in the
same manner as the data, and summing over each halo in the stellar
mass bin, weighting each galaxy by the bias value $b_h(\mhalo)$ given
by the \cite{tinker_etal:10_bias} bias function. For sub-halos, we use
the mass of the host halo (the halo that the subhalo is located
within) to calculate the bias. More explicitly, the bias of a sample
galaxies is given by summing over each galaxy, $i$, in the sample:

\begin{equation}
b_{\rm gal} = \frac{1}{N_{\rm gal}}\sum_i b_h(M_{\rm host}^{(i)}),
\end{equation}

\noindent where $N_{\rm gal}$ is the total number of galaxies in the
sample. This approach allows us to calculate the large-scale bias
rapidly, without explicitly measuring the clustering of the mock
galaxies. This is also less noisy than measuring the clustering, an
important feature in this method as the BOSS data is actually a larger
volume than the MultiDark simulation.

\begin{deluxetable*}{lllll}
%\tabletypesize{\scriptsize} 
\tablecolumns{5} 
\tablewidth{30pc} 
\tablecaption{Quantities to rank-order the CMASS galaxies\label{t.props}} 
\tablehead{\colhead{Number} &\colhead{Property}
  &\colhead{code} & \colhead{IMF} &\colhead{comments}}
\startdata

1 & Stellar mass & PCA & Kroupa & BC03 SPS \\
2 & Stellar mass & PCA & Kroupa & M11 SPS \\
3 & Stellar mass & Granada & Salpeter & wide formation times, no dust \\
4 & Stellar mass & Granada & Salpeter & wide formation times, dust \\
5 & Stellar mass & Granada & Salpeter & early formation times, no dust \\
6 & Stellar mass & Granada & Salpeter & early formation times, dust \\
7 & Stellar mass & Granada & Kroupa & wide formation times, no dust \\
8 & Stellar mass & Granada & Kroupa & wide formation times, dust \\
9 & Stellar mass & Granada & Kroupa & early formation times, no dust \\
10 & Stellar mass & Granada & Kroupa & early formation times, dust \\
11 & Stellar mass & Portsmouth & Kroupa & SF template \\
12 & Stellar mass & Portsmouth & Salpeter & SF template \\
13 & Stellar mass & Portsmouth & Kroupa & passive template \\
14 & Stellar mass & Portsmouth & Salpeter & passive template \\
15 & Luminosity & --- & --- & absolute, $i$-band \\
16 & Luminosity & --- & --- & absolute, $i$-band, k-corrected \\
17 & Velocity dispersion & PCA & --- & --- \\
18 & Velocity dispersion & Portsmouth & --- & --- \\

\enddata
\end{deluxetable*}

\begin{figure*}
%\epsscale{1.2} 
\plotone{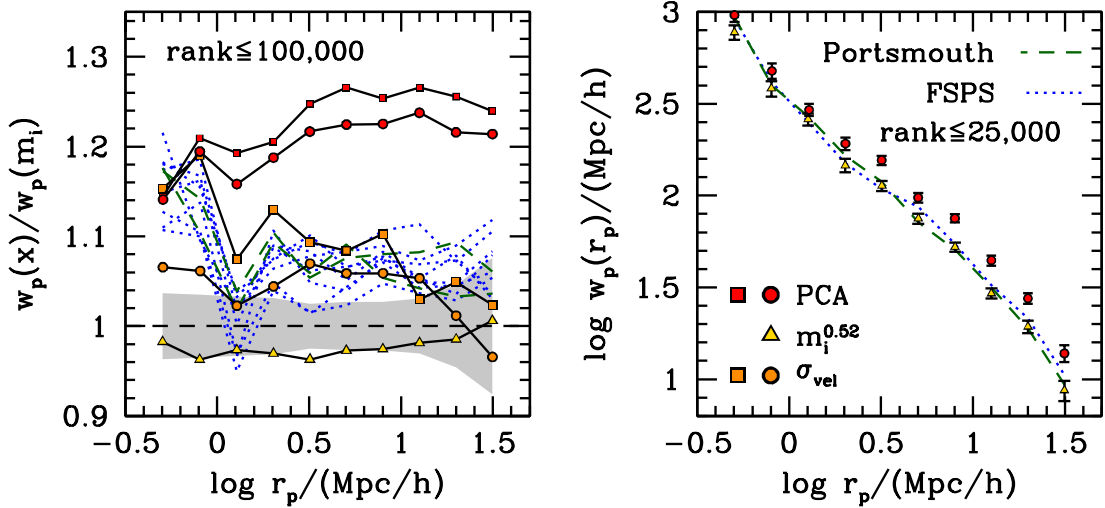}
\vspace{-8cm}
\caption{ \label{wp2} {\it Top Left}: Projected clustering relative to
clustering of BOSS CMASS galaxies ordered by $i$-band absolute
magnitude. The shaded region indicates the typical error of the
clustering measurement. The red symbols indicate the clustering when
BOSS galaxies are rank-ordered by the PCA masses. The yellow triangles
indicate clustering when ranked by $i$-band absolute magnitude
k-corrected to $z=0.52$. The blue dotted curves are the clustering
when ranked by Granada Granada masses. The green dashed curves indicate
the clustering ranked by Portsmouth masses. The orange symbols
indicate clustering when galaxies are ranked by velocity dispersion;
orange circles are the PCA estimate of velocity dispersion, while
orange squares are the Portsmouth dispersion estimate. All samples are
the top 100,000 out of 350,000 galaxies in this redshift slice. {\it
  Top Right}: Projected clustering for a subset of the samples in the
left panel (to avoid confusion). In this panel, clustering for the top
25,000 galaxies is measured for each rank-order. Red squares show the
PCA masses with BC03 SPS models. The blue dotted curve is the fiducial
Granada mass: early formation times, Kroupa IMF, dust modeling. The green
dash line is the fiducial Portsmouth mass, which uses Kroupa IMF. The
yellow triangles represent k-corrected $i$-magnitude. The differences
between PCA and other ranking methods is large enough to be seen
without taking ratios. }
\end{figure*}

\begin{figure*}
%\epsscale{1.2} 
\plotone{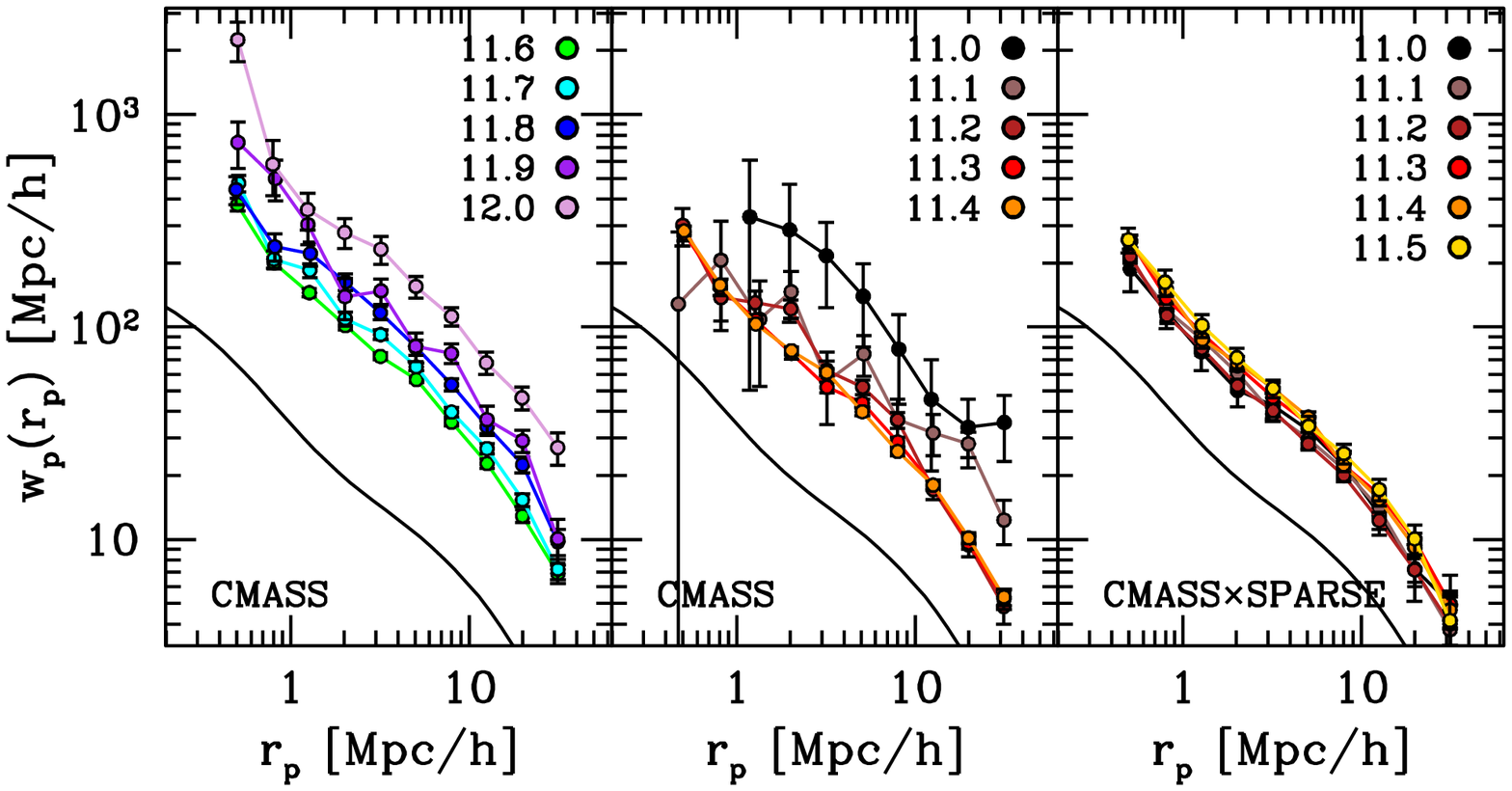}
\vspace{-6.5cm}
\caption{ \label{wp_3panel} {\it Left Panel}: Projected clustering of
  CMASS galaxies binned by stellar mass above the completeness limit
  of the CMASS selection algorithm. The solid black curve shows
  the nonlinear clustering of dark matter. {\it Middle Panel}:
  Clustering of CMASS galaxies at and below the completeness limit of
  the sample. {\it Right Panel}: Cross-correlation function of the
  SPARSE sample, binned by stellar mass, with the full CMASS
  sample. In each panel, the numbers on the right hand side indicate
  the center of the bin in $\log\mgal$.}
\end{figure*}

\section{Results}

\subsection{What galaxy property correlates most with halo mass?}

Figure \ref{wp2} compares the clustering of all 18 properties listed
in Tabel \ref{t.props} at fixed number density. For each quantity, we
rank-order the sample form highest to lowest value. In this method,
cutting the samples at a given rank means that each sample has the
same number density. In the left panel, we show $\wp$ relative to the
$i$-band $\wp$ for the top 100,000 galaxies for each property (about
$\sim 1/3$ the total sample). In the right panel, we show the results
for the top 25,000 galaxies in each property, although in the right
panel we only show a subset of the properties considered.  These two
sample sizes correspond roughly to number densities of $\sim
1.2\times10^{-4}$ and $0.3\times 10^{-4}$ $($\hmpc$)^{-3}$, respectively.

The left panel demonstrates that the observed scatter between halo
mass and luminosity is larger than all other galaxy properties
considered, regardless of the stellar mass code employed. The cabal of
photometrically-based stellar mass estimates yield a bias roughly 3\%
higher than $i$-band magnitude. Because all the different variations
of these two codes are roughly consistent with one another, we do not
attempt to differentiate them in the plot. The clustering for these
samples increases sharply at $r_p<1$ \hmpc\ relative to $i$-band
clustering. Clustering at these scales is reflective of the number of
satellite galaxies per halo (often referred to as the `one-halo'
term). At fixed luminosity, red galaxies are less massive than blue
galaxies. Because satellite galaxies are more often red than field
galaxies of the same $\mgal$, at fixed number density there will be a
higher fraction of satellites when ranked by $\mgal$ rather than by
luminosity. This argument is clearly applicable for a complete sample
like the SDSS Main galaxy sample. For the CMASS target selection,
which preferentially selects red galaxies, the impact of this argument
is less clear. However, the increase in the clustering in one-halo
term argues that the small fraction of blue galaxies in CMASS is
coming into play when rank-ordering the same by luminosity and by
stellar mass.

The amplitude of clustering for the samples defined by velocity
dispersion is consistent with the photometrically-derived stellar
masses. For some photometric stellar mass codes, the clustering of the
velocity dispersion sample is slightly higher than the photometric
stellar mass clustering, but when considering the entire ensemble of
photometric mass definitions it is not clear that there is any
differentiation between the two properties when it comes to clustering
amplitude.

The spectroscopic PCA masses have a clustering amplitude $\sim 25\%$
higher than $i$-band clustering, significantly higher than both
velocity dispersion and photometric stellar mass. Although we do not
perform any halo occupation fits of the clustering, it is improbable
to ascribe the boost in the large-scale amplitude entirely to an
increase in satellite galaxies; as opposed to the photometric stellar
masses, the relative $\wp$ for the PCA masses decreases in the
one-halo term. In the right panel, we restrict the lists to only the
most highest 25,000 objects by rank. In this sample, the difference
between the PCA clustering and the other samples is large
enough---$\sim 60\%$---to be seen clearly on a logarithmic scale, and
without taking ratios. In this panel, we show only a subset of
clustering results in order to avoid crowding of the plot.

We therefore conclude that $\slogm$, and by extension $\sigma_{\rm
  err}$, for the PCA masses is smaller than for photometrically-defined
stellar masses, galaxy luminosity, and stellar velocity
dispersion. For the remainder of this paper, all results will use the
PCA masses with the BC03 SPS code.

\subsection{Bias as a function of stellar mass}

\begin{figure}
\epsscale{1.2} 
\plotone{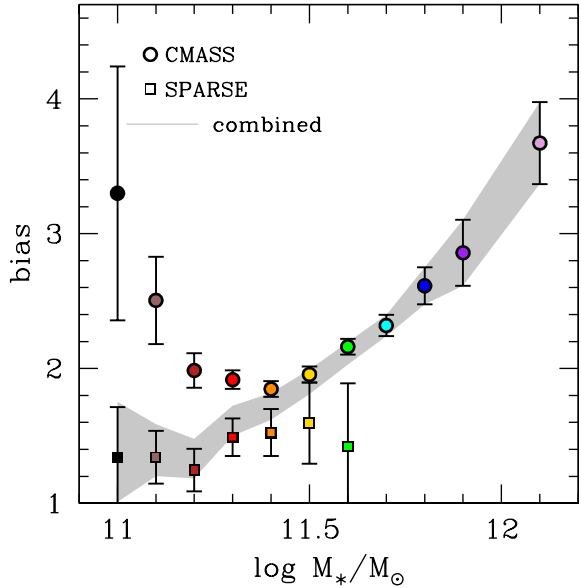}
\vspace{-0.5cm}
\caption{ \label{bias} Bias as a function of stellar mass for the
  CMASS sample (circles), the SPARSE sample (squares) and the combined
  sample (gray shaded region). }
\end{figure}

Figure \ref{wp_3panel} show the measured values of $\wp$ for CMASS and
the total CMASS sample crossed with SPARSE galaxies, binned by stellar
mass. Using the technique described in \S \ref{s.measuring_bias}, we
fit for $b_{\rm gal}$ for each bin in $\mgal$ for CMASS and for the
CMASS-SPARSE cross-correlation. Figure \ref{bias} shows the results
for each sample. The SPARSE results have had the bias of the overall
CMASS sample divided out. At $\log\mgal\ge 11.4$, the CMASS bias rises
monotonically with stellar mass. Not coincidentally, this is the mass
range where the CMASS sample is most complete in stellar mass. In
terms of space density, the CMASS sample peaks at $\log\mgal=11.4$,
and rapidly decreases at smaller masses, mainly due to the color
cuts. At $\log\mgal<11.4$, the CMASS bias rises back up again. This
rise at low masses can be explained if the target selection cuts
preferentially select satellite galaxies at lower masses. Previous
studies of color-dependent clustering have shown this U-shaped
behavior is red galaxy clustering (see, e.g.,
\citealt{swanson_etal:08_bias, ross_etal:11}), a trend that is driven by the
increasing fraction of satellite galaxies in the red subsample as
luminosity or stellar mass decreases. In contrast to those results,
the minimum bias shown in Figure \ref{bias} is at a much higher mass
relative to the knee in the stellar mass function, but it is possible
that the effect is amplified by the BOSS CMASS color cuts.

The clustering of the SPARSE sample, however, is nearly independent of
stellar mass. The \wise\ sample is not large enough to afford spatial
clustering analysis, so we make the approximation that the \wise\
clustering is the same as the SPARSE and weight the SPARSE clustering
results accordingly. The shaded band represents the total bias of
galaxies in the combined sample weighted by the relative number if the
CMASS vs. the (WISE+SPARSE) samples.

\subsection{Scatter of stellar mass at fixed halo mass}
\label{s.slogm}

Figure \ref{bias_combined} compares our combined bias results to
predictions from the numerical simulation described in \S
\ref{s.nbody}. As described previously, each curve represents a model
that matches the same stellar mass function, but has different amounts
of scatter between halo mass and stellar mass. Four different curves
representing different $\slogm$ values are shown for comparison. A
model with zero scatter would predict clustering too high relative to
the results. A model with scatter at $\slogm=0.26$ is ruled out
because the clustering amplitudes are too low. The best-fit value of
$\slogm$ is 0.18, with a 68\% confidence interval of $[0.16,0.19]$.

The SDSS pipeline reports a mean error of 0.16 dex for the PCA
masses. This includes both systematic and random errors. To estimate
$\sigma_{\rm err}$---the statistical errors alone---we use repeated
spectra of CMASS galaxies that occur on regions of the footprint
covered by multiple tiles. The rms difference in masses between the
repeat spectra is 0.11 dex, implying that the intrinsic scatter of
stellar mass at fixed $\mhalo$ is 0.16 dex. 

\subsection{The Stellar-to-Halo Mass Relation}

Figure \ref{shmr_grid} shows the stellar-to-halo mass relation (SHMR)
using the best-fit $\slogm$ value of 0.18. The solid curve represents
the mean stellar mass at fixed halo mass, $\langle \mgal |
\mhalo\rangle$. Due to the steepness of the
halo mass function, the reverse relation, $\langle \mhalo | \mgal\rangle$, is
quite different. The filled circles show the $\langle \mhalo | \mgal\rangle$
for the bins analyzed in this paper. The errors show the inner 68\% of
the distribution of $\log\mhalo$ in each bin, demonstrating the
significant overlap in the halo distribution analyzed in each bin. 

\begin{figure}
\epsscale{1.2} 
\plotone{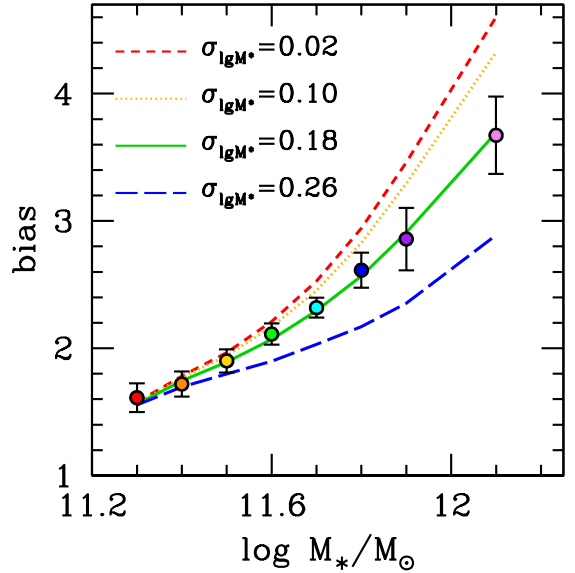}
\vspace{-0.5cm}
\caption{ \label{bias_combined} The combined CMASS+SPARSE bias values
  compared to models derived by abundance matching dark matter halos
  and subhalos in the MultiDark simulation to the stellar mass
  function of BOSS. Different curves indicate different values of
  scatter (in $\log\mgal$) in stellar mass at fixed halo mass.  }
\end{figure}

\begin{figure}
\epsscale{1.2} 
\plotone{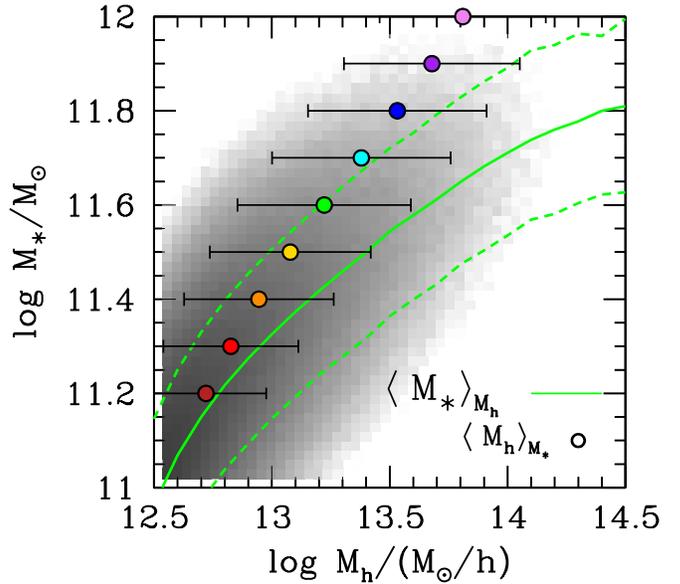}
\vspace{-1cm}
\caption{ \label{shmr_grid} The stellar-to-halo mass relation (SHMR)
  for BOSS CMASS galaxies using the PCA stellar mass estimates. The
  solid green curve shows the mean $\mgal$ in bins of $\mhalo$, with
  the dashed curves indicating the 0.18 dex in scatter of the best-fit
  relation. The colored circles show the mean $\mhalo$ in the observed
  bins in $\mgal$ used in this paper. Error bars indicate the inner
  68\% of the distribution of $\log\mhalo$ in each bin. The gray
  shaded region indicates the number of halos, scaled as $\log N_h$, at
  each point in this 2D parameter space. }
\end{figure}

\begin{figure*}
%\epsscale{1.2} 
\plotone{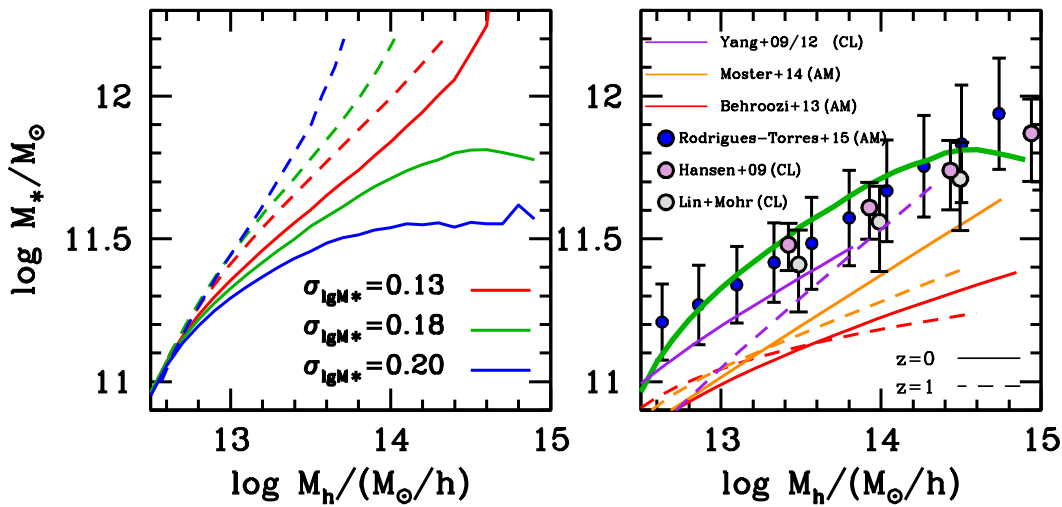}
\vspace{-7.5cm}
\caption{ \label{shmr2} {\it Left-hand side:} The sensitivity of the
  SHMR to the assumed value of $\slogm$. The solid curves show the
  SHMR for each value of scatter indicated in the key. The dashed
  curves show the mean $\mhalo$ in bins of $\mgal$. This figure
  explains why the bias in Figure \ref{bias_combined}, $b(\mgal)$,
  decreases as the scatter increases. {\it Right-hand side:}
  Comparison of the SHMR derived here to other measurements in the
  literature. There are few measurements at $z=0.5$, show we show
  values at $z=0$ and $z=1$ from the same works, with the expectation
  that the $z=0.5$ value should lie somewhere in between. Estimates of
  the SHMR from abundance matching, such as \cite{behroozi_etal:13}
  and \cite{moster_etal:13}, lie significantly below the relation
  here. Values derived from cluster samples appear to be in much
  better agreement with BOSS. Lastly, we are in good agreement with
  the results of \cite{rodriguez_torres_etal:16}, who also analyze the
  CMASS sample but using a different stellar mass estimate and
  abundance matching technique. }
\end{figure*}

The left-hand side of Figure \ref{shmr2} shows the sensitivity of the
SHMR to the assumed value of $\slogm$. At $\slogm=0.13$, the SHMR is a
steeply rising function of $\mhalo$. At $\slogm=0.20$, $\mgal$ becomes
nearly independent of halo mass at $\log\mhalo\gtrsim14$. The reverse
relations, $\langle \mhalo | \mgal\rangle$ shows the opposite trend; as
$\slogm$ gets larger, the mean halo mass at fixed $\mgal$ decreases,
yielding the trend of the theoretical models in Figure
\ref{bias_combined} and producing the tight constraints on $\slogm$.

The right-hand side of Figure \ref{shmr2} compares our best-fit
relation to a sample of other measurements. Those based purely on
abundance matching---i.e., no other data other than the stellar mass
function is fit, (\citealt{behroozi_etal:13} and
\citealt{moster_etal:13} in the Figure)---lie significantly below our
relation. This comparison is complicated by different stellar mass
estimates and significantly smaller data samples at $z=1$, but it is
clear the estimates that probe this relation directly using galaxy
groups and clusters are in better agreement with the BOSS relation
(\citealt{lin_mohr:04, hansen_etal:09, yang_etal:09_groups3,
  yang_etal:12}). Last, our results are in good agreement with the
SHMR for BOSS galaxies obtained by
\cite{rodriguez_torres_etal:16}. This relation is based on the
Portsmouth stellar masses (line 13 in Table 1) and the abundance
matching is based on peak circular velocity, with a scatter of
$0.31\times V_{\rm peak}$ obtained by matching the redshift space
monopole of the galaxy correlation function. 

\section{Discussion}

In this paper we have demonstrated a novel method of discriminating
between disparate methods of estimating the stellar masses of
galaxies. We caution that the clustering results represent only half
the story when it comes to appraising different methods because
clustering is only sensitive to the rank-ordering of a set of
galaxies, from most massive to least massive, which in turn provides
information on the scatter induced in the estimation of
$\mgal$. However, the method has nothing to say about absolute offsets
between different methods.  Using this technique, that stellar masses
derived from the PCA code of \cite{chen_etal:12} correlate stronger
with halo mass than $i$-band magnitude, velocity dispersion, and other
estimates of stellar mass provided in the BOSS pipeline. These other
stellar masses are based on photometric data only, while the PCA
masses are based on analysis of the spectral information. The PCA
method has the limitation that the spectra only contain information
about the galaxy from within the diameter of the fiber, which for BOSS
is $2\arcsec$. This may lead to aperture bias for the derived
quantities, but any bias goes in the direction of strengthening the
correlation with halo mass, which indicates something fundamental
about the quantities being estimated. But given the average radius of
the typical BOSS galaxy of $1.2\arcsec$ (\citealt{masters_etal:10}),
aperture bias might be non-negligible but it is unlikely to be the
dominant source of the differences between the methods.

\cite{bundy_etal:15} present a detailed comparison of the Porstmouth
and PCA masses with masses obtained using extra infrared imaging
available in the SDSS southern equatorial stripe, `Stripe 82' in the
collaboration parlance. The dispersion between the Portsmouth
masses\footnote{The Portsmouth mases used here, which are a
  combination of the two templates based on galaxy color, are referred
  to in \cite{bundy_etal:15} `Porstmouth Best'.} and the near-IR masses is
0.29 dex, while the dispersion between near-IR and PCA masses
0.20. The tighter correlation with near-IR masses supports the results
here that the PCA masses have a smaller intrinsic scatter relative to
other photometric-based methods. 

As discussed above, however, there is a bias between the near-IR and
PCA of 0.15 dex as well, as well as an offset of the PCA masses with
respect to the Porstmouth values
(\citealt{chen_etal:12}). Addtionally, the lower amplitude of the PCA
$\wp$ at $r_p<1$ Mpc$/h$ is indicative of fewer satellite galaxies in
the PCA sample relative to the other stellar mass samples (cf. Figure
\ref{wp2}). This does not mean that satellites are `missing' from the
PCA catalog, but rather that they are assigned lower masses than in
other catalogs. Satellites should be redder than the overall
population, implying that the PCA method finds higher stellar masses
for bluer galaxies relative to other
methods. \cite{tinker_etal:12_cosmos} found that X-ray groups with
bluer central galaxies have higher clustering at fixed halo mass, at
the same redshift and stellar mass as the BOSS
sample. \cite{tinker_etal:12_cosmos} concluded that this was assembly
bias in massive galaxies, with the caveat that the sample was
statistically limited. It is possible that the higher clustering
amplitude of the PCA masses is partly due to an assembly bias effect
imparted by the relative ranking of bluer and redder galaxies within
the catalog, and not entirely from a minimization of intrinsic
scatter. The evidence for this is circumstantial at best, but is worth
further investigation.

It is noteworthy that all stellar mass estimates considered here
correlate with halo mass better than absolute magnitude. The CMASS
sample, although it is primarily intended to target luminous red
galaxies, does have a non-negligible component of star formation
galaxies with blue(ish) colors that have a lower clustering amplitude
at fixed $M_i$ (\citealt{guo_etal:13}). We conclude that almost any
estimate of stellar mass provides a more robust rank-ordering of a
heterogeneous color sample of galaxies relative to luminosity alone.

In our analysis, the photometric stellar mass indicators provide the
same level of scatter as velocity dispersion. All of which yield
larger scatter than the PCA stellar masses. \cite{wake_etal:12} used
clustering to claim that velocity dispersion had a stronger
correlation with halo mass than stellar mass, a result that was
challenged by \cite{li_etal:13}, who claim that the Wake
et.~al.~result is contaminated by satellite galaxies, and once these
satellites are removed, stellar mass has the strongest
correlation. Our results suggest that the choice of stellar mass
estimator can play a large role in this comparison. Both Wake
et.~al.~and Li et.~al.~employ photometric mass estimates which are
likely to include extra measurement scatter, although we do not test
their exact methods. We make no attempt to remove satellite galaxies
from our analysis, but the satellite fraction of BOSS galaxies is low
($\fsat=0.10\pm0.02$ from \citealt{white_etal:11}), and it is clear in
Figure \ref{wp2} that the PCA clustering in the one-halo term
($r_p\lesssim 1 Mpc/h$) is lower relative to the photometrically-based
codes. This implies that the satellite fraction of the PCA sample is
smaller than the satellite fraction of the other samples, thus not
artificially enhancing the clustering of the sample. Our theoretical
modeling includes subhalos in the abundance matching procedure, and
this method yields the proper satellite fraction of BOSS galaxies
(\citealt{nuza_etal:13}).

Our constraint on $\slogm$ of $0.18^{+0.01}_{-0.02}$ compares
favorably with other estimates from the
literature. \cite{zu_mandelbaum:16} use abundance and lensing data to
obtain $\slogm=0.20\pm0.01$ at $\mhalo=10^{13}$ \hmsol\ (linearly
interpolating between their results at $10^{12}$ and $10^{14}$).
\cite{more_etal:11}, using satellite kinematics, find 68\% confidence
regions of $[0.14,0.21]$ for red galaxies and $[0.07,0.26]$ for blue
galaxies in the SDSS Main sample. \cite{reddick_etal:13}, using the
SHAM approach with multiple free parameters constrained by both
clustering and comparison to an SDSS group catalog, find
$\slogm=0.20\pm0.03$. \cite{leauthaud_etal:12_shmr} apply the SHMR
approach to multiple statistics in the COSMOS field to find
$\slogm=0.25\pm0.02$ at similar redshifts to those probed here. Our
constraint has both the smallest uncertainty and the lowest value
itself. The previous measurements used photometrically-derived stellar
masses, which must be contributing to the measurement scatter in each
study. \cite{kravtsov_etal:14} measure the scatter in brightest
cluster galaxy mass for a sample of X-ray clusters, findering
$\slogm=0.17\pm 0.02$, which is excellent agreement with our
results. The notable aspect of the comparison between
\cite{kravtsov_etal:14} and our results is that they are largely
distinct in the halo masses probed; Kravtsov et.~al.~analyze clusters
at $\mhalo \gtrsim 10^{14}$ \hmsol, which is the upper limit of the
halo masses probed by out stellar mass bins. This implies that
$\slogm$ is independent of halo mass over the range $12.7\lesssim
\log\mhalo \lesssim 15.2$. This has largely been assumed, mainly
because existing data could be fit with a constant scatter, and
constraints on $\slogm$ at low halo and galaxy masses are very weak
because of the lack of variation of bias with halo mass at those
scales. Additionally, all of these results imply little to no redshift
evolution in $\slogm$ for massive galaxies. This is expected, given
that the maority of the galaxy population at these masses is passively
evolving, although strict passive evolution does not fit the evolutino
of the clustering of massive galaxies over the same timespan
(\cite{zhai_etal:16}).

\cite{gu_etal:16} investigate the origin of scatter at fixed halo mass
by following the hierarchical buildup of both dark and stellar mass in
simulations using abundance matching as a function of cosmic
time. They find that $\slogm$ from merging alone (e.g., `ex-situ'
stellar mass growth) can account for 0.16 dex of scatter at
cluster-scale halo masses. This is in excellent agreement with
\cite{kravtsov_etal:14}, but the comparison to the BOSS results is
more nuanced. Our SHMR indicates that the average $10^{14}$ \hmsol\
halo contains at $10^{11.5}$ $\msol$ galaxy, but once binned by
stellar mass, the average halo at that mass scale is $\sim 10^{13}$
\hmsol. The \cite{gu_etal:16} simulations indicate that, at that halo
mass, `in-situ' processes dominate. The abundance matching analyses
indicate that the fraction of stellar mass from merging for
$\mhalo\sim 10^{13}$ \hmsol\ is low, $\lesssim 10\%$ from
\cite{moster_etal:13} and $\lesssim 25\%$ from
\cite{behroozi_etal:13}. Thus CMASS galaxies, although they are among
the most massive in the universe, still are a sensitive probe of the
physics of galaxy formation.

%We have used the subhalos identified in the MultiDark simulation to
%account for the effect of satellite galaxies in our analysis, making
%the implicit assumption that the relation between stellar mass and
%halo mass is the same for halos and subhalos. However, the bias of a
%galaxy sample is influenced by the fraction of galaxies that are
%satellites because they occupy highly biased, massive halos. Figure
%\ref{bias_fsat} shows that our results have little sensitivity to the
%satellite fraction of BOSS galaxies. \cite{white_etal:11} constrained
%the satellite fraction of CMASS galaxies to be $\fsat=0.10\pm0.02$,
%but in order to significantly alter the $b(\mgal)$ relation, we
%require a factor of $3-5$ increase in $\fsat$ over that
%constraint. Marginalizing over the 2\% constraint on $\fsat$ in White
%et.~al.~ would have negligible effect on the $\slogm$ constraints.

\acknowledgements IZ is supported by NSF grant AST-1612085.  Funding
for the Sloan Digital Sky Survey IV has been provided by the Alfred
P. Sloan Foundation, the U.S. Department of Energy Office of Science,
and the Participating Institutions. SDSS-IV acknowledges support and
resources from the Center for High-Performance Computing at the
University of Utah. The SDSS web site is www.sdss.org.

SDSS-IV is managed by the Astrophysical Research Consortium for the
Participating Institutions of the SDSS Collaboration including the
Brazilian Participation Group, the Carnegie Institution for Science,
Carnegie Mellon University, the Chilean Participation Group, the
French Participation Group, Harvard-Smithsonian Center for
Astrophysics, Instituto de Astrof\'isica de Canarias, The Johns
Hopkins University, Kavli Institute for the Physics and Mathematics of
the Universe (IPMU) / University of Tokyo, Lawrence Berkeley National
Laboratory, Leibniz Institut f\"ur Astrophysik Potsdam (AIP),
Max-Planck-Institut f\"ur Astronomie (MPIA Heidelberg),
Max-Planck-Institut f\"ur Astrophysik (MPA Garching),
Max-Planck-Institut f\"ur Extraterrestrische Physik (MPE), National
Astronomical Observatory of China, New Mexico State University, New
York University, University of Notre Dame, Observat\'ario Nacional /
MCTI, The Ohio State University, Pennsylvania State University,
Shanghai Astronomical Observatory, United Kingdom Participation Group,
Universidad Nacional Aut\'onoma de M\'exico, University of Arizona,
University of Colorado Boulder, University of Oxford, University of
Portsmouth, University of Utah, University of Virginia, University of
Washington, University of Wisconsin, Vanderbilt University, and Yale
University.

%%%%%%%%%%%%%%%%%%%%%%%%%%%%%%%%%%%%%%%%%%%%%%%%%%%%%%%%%%%%%%%%%%%%%%%%
%  Bibliography
%%%%%%%%%%%%%%%%%%%%%%%%%%%%%%%%%%%%%%%%%%%%%%%%%%%%%%%%%%%%%%%%%%%%%%%%

%\break
\bibliography{../../risa}

\end{document}